\documentclass{aa}

\usepackage{graphics}

\begin{document}

\font\small=cmr8           
\font\petit=cmcsc10        

\def\eqd{\buildrel \rm d \over =}    
\def\p{\partial}           
\def\px{\partial _x}           
\def\py{\partial _y}           
\def\pz{\partial _z}           
\def\pt{\partial _t}           
\def\ssum{\textstyle\sum}
\def\arr{\rightarrow}
\def\id{\equiv}
\def\eqv{\leftrightarrow}
\def\fol{\rightarrow}
\let\prop=\sim
\def\gapprox{\;\rlap{\lower 2.5pt            
 \hbox{$\sim$}}\raise 1.5pt\hbox{$>$}\;}       
\def\lapprox{\;\rlap{\lower 2.5pt            
 \hbox{$\sim$}}\raise 1.5pt\hbox{$<$}\;} 

\def\ang{\,{\rm\AA}}
\def\cm{\,{\rm cm}}
\def\km{\,{\rm km}}
\def\kpc{\,{\rm kpc}}
\def\second{\,{\rm sec}}     
\def\erg{\,{\rm erg}}
\def\ev{\,{\rm e\kern-.1em V}}
\def\kev{\,{\rm ke\kern-.1em V}}
\def\k{\,{\rm K}}
\def\K{\,{\rm K}}
\def\gauss{\,{\rm gauss}}
\def\SFU{\,{\rm SFU}}

\def\R{\,{\rm I\kern-.15em R}}
\def\N{\,{\rm I\kern-.15em N}}

\def\mhz{\,{\rm MHz}}

\def\n{\noindent}
\def\lead{\leaders\hbox to 10pt{\hfill.\hfill}\hfill}
\def\a{\"a}
\def\o{\"o}
\def\u{\"u}
\def\infinit{\infty}
\def\upr#1{\rm#1}

\thesaurus{09 (06.06.3; 02.13.2;  02.03.1; 02.20.1; 03.13.3)}

\title{MHD consistent cellular automata (CA) models I:
Basic features}

\author{H.\ Isliker \inst{1} \and A.\ Anastasiadis \inst{2} \and 
L.\ Vlahos \inst{1}}

\institute{
Section of Astrophysics, Astronomy and Mechanics \\
Department of Physics, University of Thessaloniki \\
GR 54006 Thessaloniki, GREECE \\
isliker@helios.astro.auth.gr, vlahos@helios.astro.auth.gr 
\and
Institute for Space Applications and Remote Sensing \hfill\break
National Observatory of Athens  \hfill\break
GR 15236 Penteli, GREECE   \hfill\break
anastasi@space.noa.gr \hfill\break}

\date{Received ...; accepted ...}

\maketitle

\begin{abstract}
A set-up is introduced which can be superimposed onto the existing 
solar flare cellular automata (CA) models, and which specifies the 
interpretation of the model's variables. It extends the CA models, yielding 
the magnetic field, the current,
and an approximation to the electric field, 
in a way that is consistent with Maxwell's and the MHD equations.
Applications to several solar flare CA models during their natural state 
(self-organized criticality (SOC)) show, among others, that (1) the  
magnetic field exhibits {\it characteristic large-scale organization} over the 
entire modeled volume; (2) the magnitude of the 
current seems spatially dis-organized, with no obvious tendency towards 
large-scale structures or even local organization;
(3) bursts occur at sites with increased current, and after a burst the current
is relaxed;
(4) by estimating the energy released in individual bursts with the use of 
the current as Ohmic dissipation, it turns out that the power-law 
distributions of the released energy persist. 
The CA models, extended with the set-up, can thus be considered as 
{\it models for energy-release through current-dissipation}.
The concepts of power-law loading and anisotropic
events (bursts) in CA models are generalized to 3--D vector-field models, 
and their effect on the magnetic field topology is demonstrated.

\end{abstract}

\section{Introduction}

\n Cellular automata (CA) models for solar flares are successful in explaining
solar flare statistics (peak flux, total flux, and duration distributions;
Lu \& Hamilton 1991 (hereafter LH91); Lu et al.\ 1993; Vlahos et al.\ 1995; 
Georgoulis \& Vlahos 1996, 1998; Galsgaard 1996).
They simplify strongly the details of the involved physical processes, and 
achieve in this way to model large volumes with complex field topologies and 
a large number of events. On the other hand, MHD simulations give insight into 
the details of the local processes, they are limited, however, to modeling 
relatively small fractions of active regions, due to the lack of computing
power, yielding thus poor statistics and difficulties in comparing results to 
observations (e.g.\ 
Mikic et al.\ 1989;
Strauss 1993;
Longcope \& Sudan 1994;
Einaudi et al.\ 1996;
Galsgaard \& Nordlund 1996;
Hendrix \& Van Hoven 1996;
Dmitruk \& Gomez 1998;
Galtier \& Pouquet 1998;
Georgoulis et al.\ 1998; 
Karpen et al.\ 1998;
Einaudi \& Velli 1999). The {\it global} MHD flare 
models are still in the state of rather qualitative flare scenarios.

The MHD and the CA approach to solar flares seem to have very little in 
common: The former are a set of partial differential equations, based on
fluid-theory and Maxwell's equations, whereas the latter are a set of
abstract evolution rules, based (in the case of solar flares) on the analogy
to critical phenomena in (theoretical) sand-piles.
The scope of this paper is to bridge the gap in-between these two approaches: 
the solar flare CA models are re-interpreted and extended so as (i) to make 
these models completely compatible with MHD and with Maxwell's 
equations, and so that (ii) all relevant MHD variables are made available 
(e.g.\ the current and the electric field, which so far were not available in 
CA models). 

In an earlier paper (Isliker et al.\ 1998), we have analyzed the existing 
solar flare CA models for their soundness with MHD. We asked the
question whether the fields in these CA models and the evolution rules can be 
interpreted in terms of MHD. It turned out that these models can indeed be 
interpreted as a particular way of implementing numerically the MHD equations. 
This fact is not trivial, since these models had been derived 
in quite close analogy to the sand-pile CA model of Bak et al.\ (1987 and 
1988), with vague association of the model's variables with physical 
quantities.
For instance, some authors (Lu et al.\  1993) explicitly discuss 
the question whether their basic grid variable is the magnetic field or not, 
without reaching to a definite conclusion. Isliker et al.\ (1998) brought 
forth not only how the existing CA models are related to MHD and what 
simplifications are hidden, but also where they differ from or even
violate the laws of MHD and Maxwell's equation. Important
is the fact that though the existing CA models can be considered as a strongly 
simplified numerical solution of the (simplified) MHD equations, {\it they do not 
represent the discretized MHD equations}: the time-step and the spacing
between two grid sites are not small (in a physical sense), but
finite; they are a typical temporal and spatial scale of the diffusive 
processes involved (see Isliker et al.\ 1998).

From the point of view of MHD, the main short-comings of the existing CA 
models are (Isliker et al.\ 1998): (1) There is no control over consistency 
with Maxwell's equations. 
Interpreting, for instance, the vector-field in the CA models as the magnetic 
field leads to the problem that the gradient of the field ($\nabla \vec B$) 
cannot be controlled. (2) Secondary quantities, such as currents, are not 
available, and they cannot be introduced in the straightforward way by 
replacing
differential expressions by difference-expressions, since, as mentioned, the 
grid-size must be considered finite (see also App.\ B.1). This lack of knowing 
how to calculate derivatives made it also useless to interpret the primary 
vector-field in the CA models
as the vector potential (to avoid the $\nabla \vec B$-problem), since $\vec B$
could not be derived. The physical interpretation of these CA models remained 
so far problematic. 

There are two basically different ways of developing CA models for flares 
further: (i) Either one considers CA models {\it per se}, tries to change the 
existing models further or invent new ones, with the only aim of adjusting 
them to reproduce still better the observations, i.e.\ one  makes them a tool
the results of which explain and maybe predict observed properties of flares.
In this approach, one has not to care about possible inconsistencies with 
MHD or even Maxwell's equations, the various components of the model are purely
instrumentalistic. (ii) On the other hand, one may care about the
physical identification and interpretation of the various components of the 
model, 
not just of its results, and one may want the CA model to become consistent
with the other approach to solar flares, namely MHD. In the 
approach (ii), some of the freedom one has in constructing CA models will 
possibly be reduced, since there are more 'boundary conditions' to be 
fulfilled in the construction of the model: the observations must be 
reproduced and consistency with MHD has to be reached. 
(Trials to construct new CA models which are based on MHD and not on the
sand-pile analogy were recently
made by 
Einaudi \& Velli (1999), MacPherson \& MacKinnon (1999), Longcope and 
Noonan (2000), and Isliker et al.\ (2000a).)

Our aim is in-between these two alternatives: we construct a set-up which can
be superimposed onto each classical solar flare CA model, and which makes 
the latter interpretable in a MHD-consistent way 
(by {\it classical} CA models we mean the models of LH91, Lu et al.\ 1993, 
Vlahos et al.\ 1995, Georgoulis \& Vlahos 1996, 1998, Galsgaard 1996, and 
their modifications, which are based on the sand-pile analogy). The set-up 
thus specifies the physical 
interpretation of the grid-variables and allows the derivation of quantities 
such as currents etc. It does not interfere with the dynamics of the CA 
(unless wished): loading, redistributing (bursting), and the appearance of 
avalanches and self-organized criticality (SOC), if the latter are implied by 
the evolution rules, remain unchanged. The result is therefore still a CA 
model, with all the advantages of CA, namely that they are fast, that they 
model large spatial regions (and large events), and therewith that 
they yield
good statistics. Since the set-up introduces all the relevant physical 
variables
into the context of the CA models, it automatically leads to a better physical 
understanding of the CA models. It reveals which relevant plasma processes 
and in what form are actually implemented, and what 
the global flare scenario is the CA models imply. All this was more or less 
hidden so far in the abstract evolution rules. It leads also to the 
possibility to change the CA models (the rules) at the guide-line of MHD, if 
this should become desirable. Not least, the set-up opens a way for further 
comparison of the CA models to observations.

In Sec.\ 2, we introduce our set-up. 
Applying it to several CA models (Sec.\ 3), we will demonstrate 
the usefulness and
some of the benefits such extended models (i.e.\ classical models extended with
our set-up) provide over the classical CA models, and we will
reveal basic physical features of the CA models.
The potential of the extended models to explain more observational facts than 
the classical CA models is, among others, outlined in the conclusions 
(Sec.\ 4).

\section{Introduction of the set-up}

\n
The set-up we propose can be superimposed onto solar flare 
CA models which use a 3--D grid and a basic 3--D vector grid-variable, say 
$\vec A$. The corresponding set of evolution rules is not changed.
(With a few modifications, the set-up can also be superimposed onto 
CA models which use a scalar field in a planar grid, which 
our set-up necessarily interprets as slab geometry, as will become
clear later.)  

We introduce our model on the example of the solar flare CA model of LH91,
which we summarize here in order to make the subsequent presentation 
more concrete:

\subsection{Summary of the CA model of LH91}

\n
In the LH91 model, to each grid-site $\vec x_{ijk}$ of a 3--D cubic grid a 
3--D vector $\vec A_{ijk}$ is assigned. Initially, $\vec A_{ijk}$ is set to 
$(1,1,1)^T$, everywhere. 
The system is then loaded with the repeated dropping of increments
at randomly chosen sites $\vec x_{ijk}$ (one per time-step)
\begin{equation}
\vec A(t+1,\vec x_{ijk}) = \vec A(t,\vec x_{ijk}) + \delta \vec 
A(t,\vec x_{ijk}),
\end{equation}
where $\delta \vec A(t,\vec x_{ijk})$ has all its components as random numbers 
uniformly distributed in $[-0.2,0.8]$.

After each loading event, the system is checked for whether the local 
'stress', defined as 
\begin{equation}
d\vec A_{ijk} := \vec A_{ijk}-{1\over 6} \sum\limits_{n.n.} \vec A_{n.n.},
\end{equation}
where the sum goes over the six nearest neighbours of the central point 
$\vec x_{ijk}$, exceeds a threshold $A_{cr}$, i.e.\ whether 
\begin{equation}
\left\vert d\vec A_{ijk} \right\vert > A_{cr},
\end{equation}
where $A_{cr} = 7$ is used. If this is the case, the field in the neighbourhood
of the critical site is redistributed according to
\begin{equation}
\vec A_{ijk}  \longrightarrow  \vec A_{ijk} - {6\over 7} d\vec A_{ijk} \\
\end{equation}
for the central point, and
\begin{equation}
\vec A_{n.n.} \longrightarrow  \vec A_{n.n.} + {1\over 7} d\vec A_{ijk} 
\end{equation}
for its six nearest neighbours. 
In such a redistribution event (burst), energy 
 amounting to
\begin{equation}
E_{rel.}^{\rm LH91} = {6\over 7} \left\vert d\vec A_{ijk} \right\vert ^2
\end{equation}
is assumed to be released.
The grid is scanned again and again to search for second, third etc.\ 
generation bursts, until the system is nowhere critical anymore and returns 
to the loading phase (the details we apply concerning the temporal evolution 
of the model are given in App.\ A; they are not explicitly stated 
in LH91).
The field outside the grid is held constant and assumed to be zero.

\subsection{Our set-up}

\n
We turn now to introducing our set-up, starting with a specification:
We interpret the vector $\vec A_{ijk}$ at the grid sites $\vec x_{ijk}$ to denote 
the local vector-field, $\vec A(\vec x_{ijk})$. Note that this was not specified 
in the classical CA models. Lu et al.\ (1993) for instance discuss this 
point: it might also have been thought of as a mean local field, i.e. the 
average over an elementary cell in the grid.

Guided by the idea that we want to assure $\nabla \vec B = 0$ for the 
magnetic field $\vec B$, which is most easily achieved by having the 
vector-potential $\vec A$ as the primary variable and letting $\vec B$ be the 
corresponding derivative of $\vec A$ ($\vec B = \nabla\wedge \vec A$), we 
furthermore assume that the grid variable $\vec A$ of the CA model is 
identical with the vector-potential.

The remaining and actually most basic problem then is to find an adequate 
way to calculate
derivatives in the grid. In general, CA models assume that the grid-spacing is
finite, which also holds for the CA model of LH91 (as shown in detail by 
Isliker et al.\ 1998),
so that the most straightforward way of replacing differential expressions
with difference expressions is not adequate (see the detailed discussion in 
App.\ B.1, below; Vassiliadis et al.\ (1998) suggested to interpret CA models
as the straightforwardly discretized (simplified) MHD equations, which we
find problematic for the reasons given in App.\ B.1, and we therefore do 
not follow this approach, here). 

Consequently, one has to find a way of continuing the vector-field
into the space in-between the grid-sites, which will allow to calculate
derivatives. 
There is, of course, an infinite number of possibilities to do so, 
and the problem cannot have a unique solution. Adequate possibilities 
definitely include:
a) continuation of $\vec A$ with the help of an equation (e.g.\ demanding the 
resulting $\vec B$-field to be potential or force-free); b) interpolation, 
either locally (in a neighbourhood), or globally (through the whole grid).
Trying several methods,
we concluded that 3--D cubic spline interpolation is 
particularly adequate to the problem since it has remarkable advantages over 
other methods (e.g.\ it does not introduce oscillations in-between grid-sites, 
which
would strongly influence the values of the derivatives, and it well reproduces 
the derivatives of analytically prescribed primary fields).
The process of evaluating different continuation methods we went through, as 
well as the comparison of spline interpolation to other continuation-methods
are described in App.\ B.

The 3--D interpolation is performed as three subsequent 1--D interpolations in
the three spatial directions (Press et al.\ 1992).
For the 1--D splines, we assume natural boundaries (the second derivatives are
zero at the boundaries). Moreover, since in the CA model of LH91 it is assumed 
that around the grid there is a zero field which is held constant (see Sec.\ 
2.1), 
we enlarge the grid by one grid point in all directions to include this 
constant zero-layer explicitly, using it however only for the 
interpolation. In the interpolation, the derivatives at the grid-points are 
immediately given by the analytically differentiated interpolating 
polynomials. 

With the help of this interpolation, the magnetic field $\vec B$ and the 
current $\vec J$ are calculated as derivatives of $\vec A$, according to the 
MHD prescription:
\begin{equation}
\vec B = \nabla \wedge \vec A,
\end{equation}
\begin{equation}
\vec J = {c \over 4\pi} \, \nabla \wedge \vec B.
\end{equation}

To determine the electric field $\vec E$, we make the assumption that 
under coronal conditions the MHD approach is in general valid, and that
$\vec E$ is reasonably well approximated by Ohm's law in its simple
form,
$\vec E = \eta \vec J - {1\over c} \vec v \wedge \vec B$, with $\eta$ the diffusivity
and $\vec v$ the fluid velocity. Since the classical CA models use no 
velocity-field, our set-up can yield only the resistive part,
\begin{equation}
\vec E = \eta \vec J.
\end{equation}
In applications such as to solar flares, where the interest is in current 
dissipation events, i.e.\ in events where $\eta$ and $\vec J$ are strongly
increased, Eq.\ (9) can be expected to be a good approximation to the 
electric field. Theoretically, the convective term in Ohm's law would in 
general yield just a low-intensity, background electric field.

Eq.\ (9) needs to be supplemented with a specification of the diffusivity 
$\eta$: Isliker et al.\ (1998) have shown that in the classical 
CA models the diffusivity adopts the values $\eta =1$ at the unstable 
(bursting) sites, and $\eta =0$ everywhere else. This specifies 
Eq.\ (9) completely.

\medskip
\noindent
{\it Remark 1:} It is worthwhile noting that, since spline interpolation has the 
property to be the least curved 
of all 
twice differentiable interpolating functions,
the grid-size is a typical smallest-possible length-scale 
of field structures, or, if we would think the CA to model 
MHD turbulence, it is something like an average smallest possible 
eddy-size. 

\medskip
\noindent
{\it Remark 2:} 
With our set-up, field lines are made available: the interpolation
was introduced to define derivatives at the grid-sites, but it can as
well be used to determine the vector-potential, and therewith the magnetic 
field and the current, in between the grid sites.
However, it is important to note that this field in between the grid sites 
is not used for the time-evolution of the model, it merely allows to visualize 
the evolution in the standard way through field-lines, if wished so
(second order derivatives in-between the grid-sites, if 
needed, would have to be done numerically, since else their numerical values 
would depend on the order in which the three 1--D interpolations are done).

\medskip
\noindent
{\it Remark 3:} 
In this paper, we do not change the rules of the classical CA models to which
we apply our set-up --- except for the definition of energy release 
(Sec.\ 3.1.3). Our aim here is to show the usefulness of the set-up and to 
give some results and reveal some aspects 
by extending published, classical CA models.
While the redistribution rules have in detail been shown to represent 
diffusion events and fit nicely into an MHD scenario (Isliker et al.\ 1998), 
the loading process is strongly simplified and poorly follows a reasonable 
flare scenario:
For instance, the loading process acts independently everywhere in the 
simulation box, whereas according to a realistic flare scenario (see e.g.\ 
Parker 1993) disturbances should appear independently only 
on one boundary  (the photosphere, due to random foot-point motions or 
newly appearing flux), and propagate then into the interior of the simulation 
box along the magnetic field lines.

To translate such a realistic loading scenario into the language of CA models
has not been undertaken, so-far.
We just note that it would be quite straightforward to introduce a velocity 
field into the CA models: e.g.\ Isliker et al.\ (2000a) propose a CA model 
which uses a velocity field for the loading phase, but this model does not 
fall into the category of classical CA models since it does not follow the 
sand-pile analogy and uses different, MHD based, evolution rules. We leave the
problem of introducing a velocity field and a more physical loading 
process into the classical CA models for a future study.
In Isliker et al.\ 2000b, we will --- among others --- analyze in details 
what this simplified loading process physically represents.

\section{Applications of our set-up}

\subsection{Application to the CA model of Lu \& Hamilton (1991)}

Our first application is to the CA model of LH91 (see Sec.\ 2.1).
The LH91 model has a fairly long transient phase and reaches 
finally a stationary state, the so-called SOC (self-organized criticality) 
state, in which spatially spreading series of bursts (avalanches) appear, 
alternating
with quiet loading phases. The LH91 model gives basically three results 
concerning flare statistics, namely the distributions of total energy, 
peak-flux and durations, which are all power-laws with slopes that are in 
good agreement with the observations (Lu et al. 1993, Bromund et al. 1995). 

Superimposing our MHD-frame onto the LH91 model such as it stands does
not change anyone of the three results, since at this first stage we are not
interfering with the dynamics (i.e.\ the evolution rules). The set-up allows, 
however, to address several questions in MHD language: 
Our main aim in the subsequent applications is to 
demonstrate that the set-up indeed yields a new and consistent interpretation
of CA-models, to illustrate the behaviour of the secondary variables
(currents, magnetic fields), and to reveal major features of them. 
(In the subsequent runs, we use a grid of size $30\times 30\times 30$, as 
LH91 did to derive their main results.)

\begin{figure}
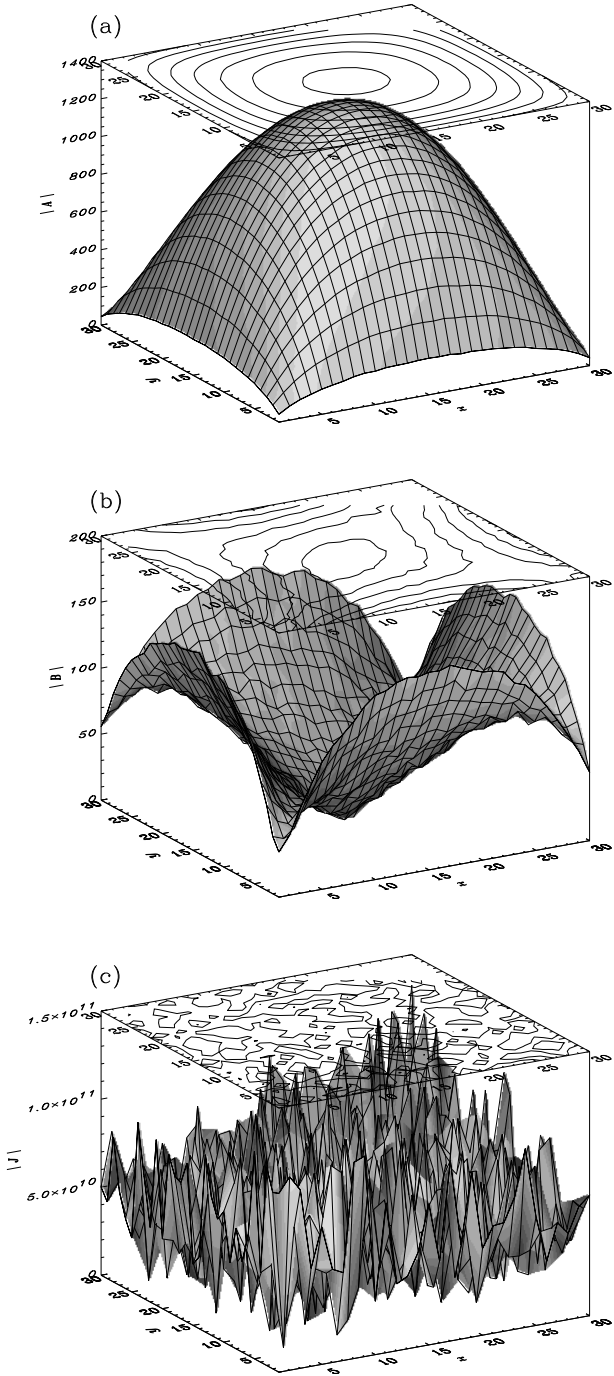

\resizebox{\hsize}{!}{\includegraphics{10131.f1a}}
\resizebox{\hsize}{!}{\includegraphics{10131.f1b}}
\resizebox{\hsize}{!}{\includegraphics{10131.f1c}}
\caption{Surface and contour plots of the magnitudes of (a) the 
vector-potential ($\vert\vec A\vert$), (b) the magnetic field 
($\vert\vec B\vert$), and (c) the current ($\vert\vec J\vert$) as a function
of $x$ and $y$, for $z=15$ fixed.}
\label{}
\end{figure}

\subsubsection{Global structures of the vector-fields}

First, we turn to the question what the global fields (vector-potential, 
magnetic field, current) look like during the SOC state. Thereto, the temporal 
evolution of the model is stopped at an arbitrary time during SOC state 
(in a phase where there are no bursts, i.e.\ during loading), and the 
magnitude of the fields at a cut with fixed $z$-coordinate are shown as a 
function of the $x$- and $y$-coordinates in Fig.\ 1. $\vert \vec A(x,y,z=z_0)\vert$  
obviously exhibits a large-scale organization over the whole grid,
it forms a global convex surface (Fig.\ 1(a)). This convex surface has
a slight random distortion over-lying, which visually cannot be discerned 
in Fig.\ 1(a), but becomes visible in the plot of $\vert\vec B\vert$ (Fig.\ 1(b)), 
the curl of $\vec A$, which still exhibits large-scale organization all over 
the grid, but is clearly wiggled. Finally, $\vert \vec J\vert $ shows no 
left-overs of a large scale organization anymore, it reflects the random 
disturbances of the convexity of $\vert \vec A\vert $ (Fig.\ 1(c)).

The large-scale structures shown in Fig.\ 1 are always maintained during the 
SOC state, neither loading nor bursting (and avalanches) destroy them, 
they just 'tremble' a little when such events occur.
SOC state in the extended LH91 model thus implies large-scale 
organization of the vector-potential and the magnetic field,
in the characteristic form of Fig.\ 1.

The large-scale organization of $\vec A$
is not an artificial result of our superimposed set-up, but 
already inherent in the classical LH91 model: in the classical LH91 CA model, 
there is only one variable, the one we call here $\vec A$, whose values are not
affected by the interpolation we perform since it is the primary grid 
variable, so that Fig.\ 1(a) is true also for the classical, non-extended 
LH91 model.

The large scale structure for the primary grid-variable 
$\vert \vec A \vert$ is the result of a combined effect: The preferred 
directionality of the loading increments (see Sec.\ 2.1) tries to increase 
$\vert \vec A \vert$ throughout the grid.
The redistribution events, which already in Bak et al.\ (1987; 1988) were
termed diffusive events, and which in Isliker et al.\ (1998) were analytically 
shown to represent local, one-time-step diffusion processes, smooth out
any too strong spatial unevenness of $\vec A$, and they root the 
$\vec A$-field down to
the zero level at the open boundaries. The result is the convex surface of
Fig.\ 1(a), blown-up from below through loading, tied to the zero-level at the 
edges, and forced to a maximum curvature which is limited by the local, 
threshold dependent diffusion events.

As the SOC state, so is the large-scale structure of $\vert \vec A \vert$ 
independent of the concrete kind of loading, provided it fulfills the 
conditions that the loading increments exhibit a preferred directionality 
and are much smaller than the threshold 
(with symmetric loading, the SOC state is actually never reached,
see LH91 and Lu et al.\ (1993)). 

To make sure of the importance of the boundaries, we performed runs of the 
model with closed boundaries, and we found that neither a large-scale 
structure was developed in $\vert \vec A \vert$, nor the SOC state was 
reached.

\begin{figure}
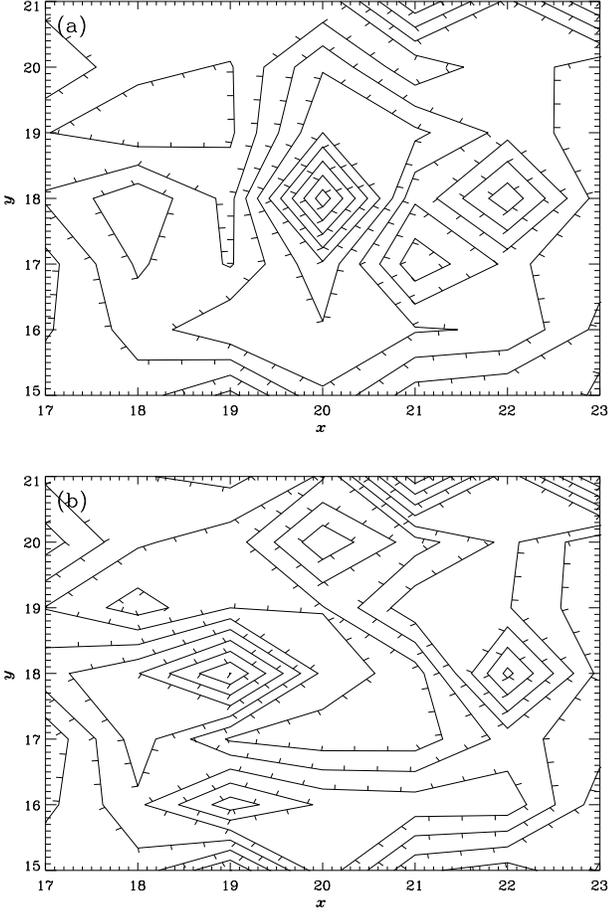

\resizebox{\hsize}{!}{\includegraphics{10131.f2a}}
\resizebox{\hsize}{!}{\includegraphics{10131.f2b}}
\caption{The magnitude of the current ($\vert\vec J\vert$, contour-plot, with the 
ticks pointing `downhill') 
as a function of $x$ and $y$ at a zoomed cut $z=3$ through the 
grid, before (a) and after (b) a burst, which occurs in the middle of the 
plot, at $(x,y,z)=(20,18,3)$.}
\label{}
\end{figure}

\subsubsection{Bursts}

To illustrate the role of the current at unstable sites and during bursts, we 
plot in Fig.\ 2 the magnitude of the current before and after a typical burst:
obviously, the current at the burst site $(x,y,z)=(20,18,3)$ has high intensity 
before the burst (Fig.\ 2(a)), and is relaxed after the burst (Fig.\ 2(b)). 
Inspecting a number of other bursts, we found that, generally, 
at sites where the LH91 instability criterion is fulfilled, 
the current is increased, too, and that bursts dissipate the currents. 
This is a first hint that classical CA models can be interpreted as models 
for energy release through current-dissipation.

After the burst, at the neighbouring site $(21,18,3)$, the intensity of the 
current is increased, and indeed the presented burst gives rise to subsequent 
bursts, it is one event during an avalanche. 

The magnetic field at the bursting site is reshaped, in a way which is 
difficult to interpret when using only the magnitude of it 
($\vert\vec B\vert$) for visualization. May-be field line plots would 
help visualization, but we leave this for a future study.

\begin{figure}
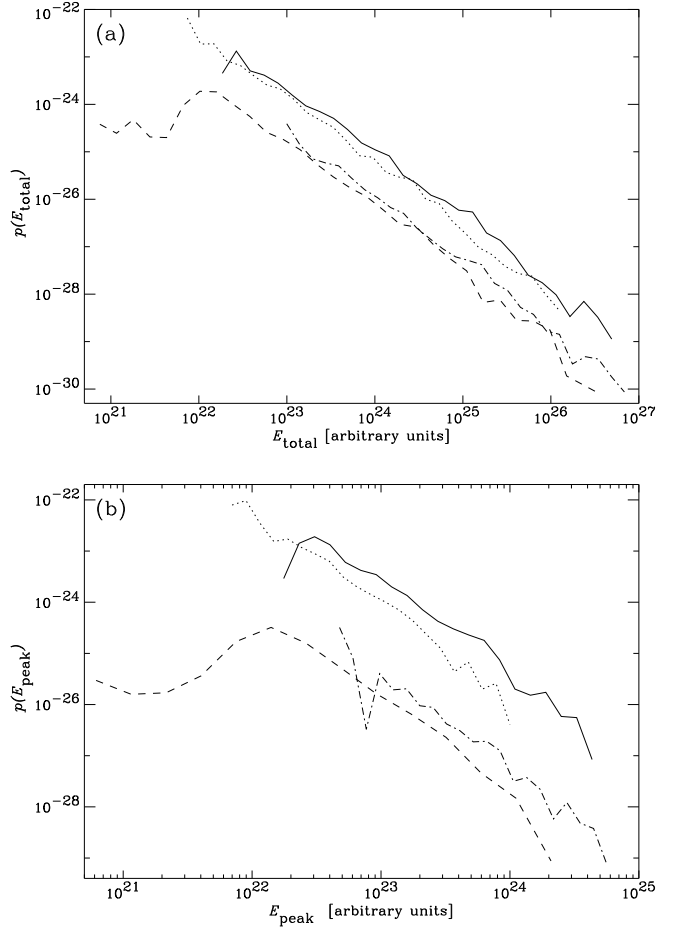

\resizebox{\hsize}{!}{\includegraphics{10131.f3a}}
\resizebox{\hsize}{!}{\includegraphics{10131.f3b}}
\caption{The distribution of the total energy (a), and of the peak-flux (b),
for different ways of measuring the released energy in a burst:
$E_{burst} = \int dt \sum_{n.n.} \vec J(\vec x_{ijk})^2$ (solid);
$E_{burst} = \int dt\, \vec J(\vec x_{ijk})^2$ (dotted);
$E_{burst}$ as the difference in 
$\sum_{n.n.} \vec B(\vec x_{ijk},t)^2$ before and after the burst (dashed);  
$E_{burst} = {6\over 7} d\vec A^2$ (dash-dotted).
(The distributions are normalized probability distributions,
the last two were shifted in both directions for viewing them together 
with the first two.)}
\label{}
\end{figure}

\subsubsection{Energy release and Ohmic dissipation}

We now turn to the question what relation the energy release formula of LH91 
(Eq.\ 6) has to the respective MHD relations: In parallel to using the 
formula of LH91, we determine the released energy in the 
following ways, closer to MHD: 
First, we assume it 
to be proportional to $\eta\vec J^2$ (with the diffusivity $\eta = 1$ at unstable
sites, see Sec.\ 2.2), which we linearly interpolate between 
the two states before and after the burst. This is done in two ways, (i)
summing over the local neighbourhood,
\begin{eqnarray}
E_{burst}^{\int\sum\vec \!\! J^2} 
&=&
\int\limits_t^{t+1} \!\!\int\limits_{n.n.} \eta \, \vec J(\vec x,t)^2 \, dt\,dV 
\approx
\int\limits_t^{t+1} \!\!\sum\limits_{n.n.} \vec J_{n.n.}(t)^2 \, dt      \cr
&=&
\sum\limits_{n.n.} {1\over 2} 
          \left(\vec J^2_{n.n.;\,before} + \vec J^2_{n.n.;\,after} \right)
\end{eqnarray}
and (ii) without summing, but just taking into account the current at the
central point,
\begin{eqnarray}
E_{burst}^{\int\!\! \vec J^2} 
&=& 
\int dt\, \eta \, \vec J(\vec x_{ijk},t)^2 \cr
&\approx& 
{1\over 2} \left(\vec J^2_{ijk;\,before} + \vec J^2_{ijk;\,after} \right)
\end{eqnarray}
and finally, we monitor the change in magnetic energy due to a burst using
the difference in magnetic energy in the local neighbourhood,
\begin{eqnarray}
E_{burst}^{\Delta \vec B^2} = 
\sum_{n.n.} && \left({\left(\vec B^{(before)}(\vec x_{n.n.})\right)^2 \over 8\pi} 
                                                       \right. \nonumber\\
 &&\left. - {\left(\vec B^{(after)}(\vec x_{n.n.})\right)^2 \over 8\pi}\right).
\end{eqnarray}
(In Eqs.\ (10), (11), (12), we assume $\Delta h = 1$ and $\Delta t = 1$ for 
the grid-spacing $\Delta h$ and the time-step $\Delta t$, since, according 
to Isliker et al.\ (1998), in the classical CA models both values are not 
specified and set to one.)

The corresponding distributions of total energy and peak-flux are shown in 
Fig.\ 3, together with the distributions yielded by the energy-release formula 
of LH91, Eq.\ (6) (the duration distribution remains the same as in the 
classical LH91 model, namely a power-law, and is not 
shown). Obviously, the four ways of defining the released
energy give basically similar results, with larger deviations only at the
low and high energy ends (note that the energy in Fig.\ 3 is in arbitrary
units). Using the formula of Ohmic dissipation does thus not change the 
results of the classical LH91 model.

With an estimate of the numerical value of the anomalous resistivity and
of the typical size of a diffusive region or the typical diffusive time, it 
would be possible to introduce physical units. We did not undertake this, 
since all three parameters are still known only with large observational and 
theoretical uncertainties.

\subsubsection{The relation of $d\vec A$ to $\vec J$}

From the similarity of the distributions of the extended model with the ones 
of the classical LH91 model (Fig.\ 3), and from Fig.\ 2, where it was seen that
an instability is accompanied by an enhanced current, we are led to
ask directly for the relation of $d\vec A$ to $\vec J$, which we plot as a 
function of each other in Fig.\ 4. Obviously, the two quantities are related 
to each other: above $\vert d\vec A\vert \approx 2$, the current is an
approximate linear function
of the stress, around $\vert d\vec A\vert \approx 2$ the current is zero,
and below there is again an approximate linear relationship, with negative 
slope, however (above $\vert d\vec A\vert \approx 2$ the current $\vec J$ is actually 
preferably along $(1,1,1)$, whereas below it is preferably along 
$(-1,-1,-1)$, i.e.\ $\vec J$ is an approximatly linear function of $d\vec A$ 
in the whole range, it merely changes its directivity at 
$\vert d\vec A\vert \approx 2$ with respect to $d\vec A$). In 
Appendix C, we show analytically why with our set-up a more or less close 
relation between $d\vec A$ and $\vec J$ has to be expected.

Of particular interest in Fig.\ 4 is that if $\vert d\vec A\vert$ is above the 
threshold $A_{cr} = 7$, then $\vert \vec J\vert$ is also reaching high values:
obviously, large values of $\vert d\vec A\vert$ imply large values of 
$\vert \vec J\vert$. This confirms the statement made above: The extended
CA models can be considered as models for energy release through current 
dissipation. It also explains why the energy distributions 
remain very similar when the LH91 formula for the amount of energy 
released in a burst ($\prop d\vec A^2$, Eq.\ (6)) is 
replaced by Ohmic dissipation ($\prop \vec J^2$, Sec.\ 3.1.3): bursts occur 
only 
for large stresses $\vert d\vec A\vert$, where $\vert \vec J\vert$ is 
also large and an approximate linear 
function of $\vert d\vec A\vert$, so that the distributions of $d\vec A^2$ and
$\vec J^2$ can be expected to be the same in shape.

\begin{figure}
\resizebox{\hsize}{!}{\includegraphics{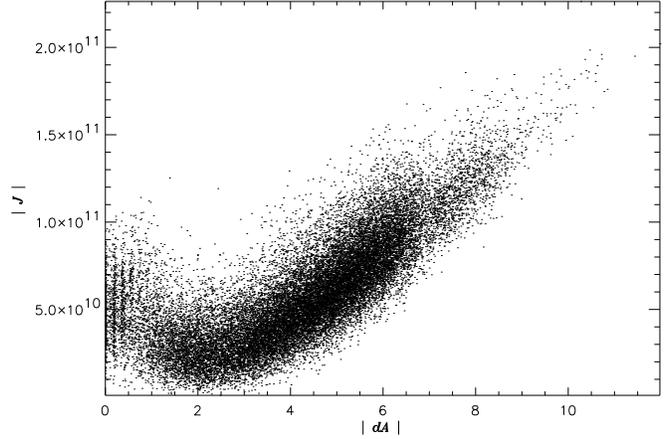}}
\caption{Plot of the magnitude of the current $\vert\vec J_{ijk}\vert$ vs.\ the
LH91 stress measure $\vert d\vec A_{ijk} \vert$, using the corresponding 
values in the whole grid at a time fixed in a loading phase in the SOC state, 
together with the values at bursting sites during an avalanche.}
\label{}
\end{figure}

\subsection{Application to loading with power-law increments}

\n Georgoulis and Vlahos (1996, 1998) introduced power-law distributed 
increments
for the loading. The main result of such a way of driving the system is that
the power-law indices of the energy-distributions depend on the power-law 
index of the 
distribution of the loading increments, explaining thus the observed 
variability 
of the indices through the variability of the intensity of the driving. We 
generalize their
way of power-law loading, which is for a scalar primary field, to a vector
field in the following way: The anisotropic directivity of the loading 
increment $\delta \vec A$ is kept (see Sec.\ 2.1), but $\vert \delta \vec A\vert$ is 
now distributed according to
\begin{equation}
p(\vert \delta \vec A\vert ) = C \vert \delta \vec A \vert ^{-\beta}
\end{equation}
with $\vert \delta \vec A\vert \in [0.01,\infty]$ and $\beta$, the power-law index, a 
free parameter. 
Simulations were performed for $\beta = 1.8$ and $\beta = 2.3$.
Interested in global features implied by the 
CA model, our concern here is the structure of the magnetic field. 
It turns out that the magnetic field exhibits still a large scale 
organization, which is very similar to the
one of the $\vec B$-field of the (extended with our set-up) LH91 model 
(Fig.\ 1(b)): for $\beta=1.8$,
the respective plots are visually indiscernible, and for $\beta=2.3$ the overall 
shape is still roughly the same, it merely seems slightly more distorted. 
Thus, though 
the statistical results depend on $\beta$, the strength and variability of the
loading, the structure of the magnetic field
remains approximately the same as in the case of the extended model of LH91.
Large-scale organization (in the characteristic form of Fig.\ 1) must 
consequently be considered as an inherent property of SOC state,
through the mechanism explained in Sec.\ 3.1.1.

\begin{figure}
\resizebox{\hsize}{!}{\includegraphics{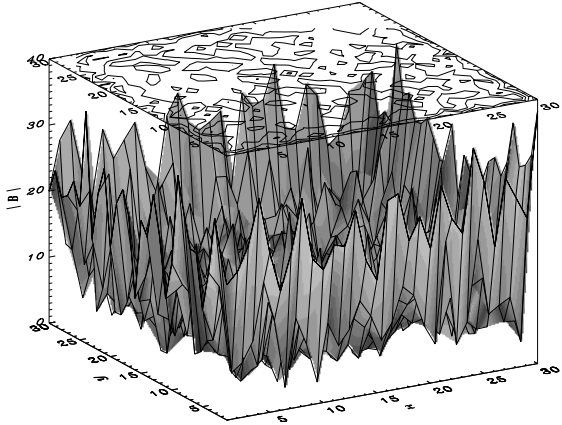}}
\caption{Surface and contour plot of the magnitude of the 
the magnetic field ($\vert\vec B\vert$) as a function
of $x$ and $y$ at a cut $z=15$ through the grid, for 
the case of anisotropic redistribution rules.}
\label{}
\end{figure}

\subsection{Application to anisotropic bursts}

\n Vlahos et al.\ (1995) introduced anisotropic bursts for solar flare CA 
models, which lead only to small events, but yield a steep distribution at 
small energies, predicting thus a significant over-abundance of small events
with a significant contribution to coronal heating. 
We have first to 
generalize the anisotropic evolution rules, which are again for a 
scalar primary field, to the case of a primary vector field.
A natural generalization would be to apply the anisotropic rules to the 
absolute magnitude of $\vec A$, but it turns out that this causes the 
algorithm to get trapped in infinite loops (two neighbouring grid-sites 
trigger each other mutually for ever). The same holds if we apply the
anisotropic rules to the absolute magnitudes of the three components of 
$\vec A$ independently. We finally applied the anisotropic rules to the three 
components of $\vec A$ directly, not using absolute magnitudes, as also 
Vlahos et al.\ (1995) did not use absolute magnitudes, and this turned out to 
lead to a stationary asymptotic state: The anisotropic stress in the 
$x$-component is thus defined as
\begin{equation}
dA^{(x)}_{ijk;n.n.} := A^{(x)}_{ijk} - A^{(x)}_{n.n.},
\end{equation}
where $n.n.$ stands for one of the six
nearest neighbours. The instability criterion is
\begin{equation}
dA^{(x)}_{ijk;n.n.} > A_{cr.}
\end{equation}
and the redistribution rules become 
\begin{equation}
A^{(x)}_{ijk} = A^{(x)}_{ijk} - {6\over 7} A_{cr.}
\end{equation}
for the central point and
\begin{equation}
A^{(x)}_{n.n.} = A^{(x)}_{n.n.} + {6\over 7} A_{cr.} 
  {dA^{(x)}_{ijk;n.n.} \over
   \sum_{n.n.}^{\prime} dA^{(x)}_{ijk;n.n.}} ,
\end{equation}
for those nearest neighbours which fulfill the instability criterion (Eq.\ 15),
where the primed sum is over those neighbours for which Eq.\ (15) holds.
The rules for $A^{(y)},A^{(z)}$ are completely analogous 
(so that actually there are 18 possibilities to exceed the threshold (Eq.\ 15)
at a given site). 
The released energy is assumed to amount to 
\begin{equation}
E_{rel}^{(aniso)} = \sum\limits_{s=x,y,z} (A^{(s)}_{ijk}-{6\over 7}A_{cr.})^2 .
\end{equation}

We performed a run where only the anisotropic burst-rules were applied, in 
order to isolate their effect, although the anisotropic burst-rules are 
used always together with the isotropic ones by Vlahos et al.\ (1995),
since alone they cannot explain the complete energy distributions of flares.
In Fig.\ 5, the magnitude of the magnetic field at a cut through the 
grid is shown (fixed $z$), for an arbitrary time (in the loading phase) during 
the asymptotic stationary state of the model. Clearly, there is no overall 
large scale structure anymore, except that
the magnetic field along the boundaries is increased. The magnetic field 
topology is thus nearer to the concept of a random, relatively unstructured 
magnetic field than the magnetic field topology yielded by the isotropic 
models in SOC state. 

The anisotropic burst rules do not yield large-scale 
structures, as they are, when used alone, also not able to lead the system to 
SOC state: this is obvious from the energy distributions they yield, 
which are much smaller in extent than the ones given by the isotropic rules 
(see Vlahos et al.\ 1995), and confirmed by the result of  Lu et al.\ (1993) 
that isotropy of the redistribution rules --- at least on the average --- 
is a prerequisite to reach SOC state, at all. 
The anisotropic bursts occur independently in all directions and 
are in this way not able to organize the field in a neighbourhood 
systematically, and, as a consequence, also not in the entire grid.

The inquiry of the relation of the energy release formula Eq.\ (18), which is 
different from the isotropic formula (Eq.\ 6), to MHD based formulae we leave 
for a future study. We just note that the distributions the anisotropic model 
in our vector-field version yields are at lower energies,
smaller in extent, and steeper than the ones of the isotropic models.

\section{Summary and Conclusions}

\subsection{Summary}

\n We have introduced a new set-up for classical solar flare CA models which 
yields, 
among others, consistency with Maxwell's equations (e.g.\ divergence-free 
magnetic field), and availability of secondary variables such as currents and 
electric fields in accordance with MHD. Both are new for solar flare CA 
models. The set-up specifies the so far open physical interpretation of the CA 
models.
This specification is to some extent unavoidably arbitrary, and it would 
definitely be interesting to see what alternative interpretations
would yield --- if they can be derived consistently. We can claim, however, 
that the interpretation we chose is 
reasonable, it is well-behaved in the sense that the derivatives of 
analytically prescribed vector-potentials are reproduced and that the 
abstract stress-measure of the CA models is 
related to the current, due to general properties of spline interpolation. The 
central problem which was to solve is how to calculate derivatives in a CA 
model, i.e.\ how to continue the primary grid-variable in-between the grid 
sites, since the notion of derivatives is alien in the context of CA models 
quite in general.

In this article, our main aim with the introduced set-up was to 
demonstrate that the set-up truly extends the classical 
CA models and makes them richer in the sense that they contain much more 
information, now. The main features we revealed about the  
CA models, extended with our set-up, are:

\noindent
{\bf 1.\ Large-scale organization of the vector-potential and the magnetic field:}
The field topology during SOC state is bound to characteristic 
large-scale structures which span the whole grid,
very pronounced for the primary grid variable, the vector-potential, but also 
for the magnetic field. Bursts and flares are just slight disturbances
propagating  over the large-scale structures, which are always maintained, 
also in the largest events. 
The magnitude of the current, as a second order derivative of the primary 
field, does not show any obvious large-scale structure anymore, it reflects 
more or less only the random fluctuations of the large-scale organized 
magnetic field. It is worthwhile noting that the large-scale structure of the 
primary grid-variable is not an artificial result of our set-up, 
but a natural consequence of the SOC state in which the system finds 
itself. 
The appearance of large-scale structures for the primary grid variable 
was shown here for the first time. It may 
have been known to different authors, but it never has explicitly been 
shown: SOC models for flares are derived in analogy to sand-pile dynamics, 
and the paradigm of a pile reappears in the field topologies of the solar 
flare CA models. 

\noindent
{\bf 2.\ Increased current at unstable grid-sites:}
Unstable sites are characterized by an enhanced 
current, which is reduced after a burst has taken place, as a result of which 
the current at a grid-site in the neighbourhood may be increased.

\noindent
{\bf 3.\ Availability of the electric field:}
The electric field is approximated 
with the resistive part of Ohm's law in its simple form,
which can in general be expected to be a good approximation in 
coronal applications and
where the interest is in current-dissipation events, e.g.\ in the case of 
solar flares.

\noindent
{\bf 4.\ Energy release in terms of Ohmic dissipation:}
We replaced the some-what {\it ad hoc} formula in the CA models 
to estimate the energy released in a burst 
with the expression for Ohmic dissipation in terms of the current. The 
distributions yielded in this way are very similar
to the ones based on the ad hoc formula, so that the results of the 
CA models remain basically unchanged. 

\noindent
{\bf 5.\ CA as models for current dissipations:}
As a consequence of point 2 and 4 in this list, and of the fact that there
is an  approximate linear relation between the current and the stress
measure of the CA, we can conclude that the {\it extended} CA models can be 
considered as models for energy release through current dissipation.

\subsection{Conclusions}

Our set-up is to be contrasted to the recently suggested MHD-derived (not 
based on the sand-pile analogy) CA models of 
Einaudi \& Velli (1999), MacPherson \& MacKinnon (1999),
Longcope and Noonan (2000), and Isliker et al.\ (2000a). They all suggest new 
evolution rules, derived from MHD, and all in different ways
(they actually focus on different processes, namely 
the microscopic, macroscopic, and mesoscopic physics, respectively, in active
regions). Our set-up, on the other hand, uses existing CA models, does
not interfere (if not wished) with their evolution rules, does also not change
their main results, as shown, but reinterprets them, extends them 
essentially, and makes them compatible with MHD.

The set-up we introduced allows different future 
applications and posing questions which could not be asked so far in the frame
of CA models. In preparation is a study (Isliker et al.\ 2000b) to reveal in 
detail
what physical flare scenario the extended CA models imply. We will 
address the questions: (1) how to interpret the small scale processes
of the models (loading and bursting) in terms of MHD;
(2) what the {\it global} flare-scenario implied by the models is;
(3) whether the global magnetic field topology of the models
can be considered to represent observed magnetic topologies in active 
regions; (4) what 
spatio-temporal evolution of the electric field during flares is yielded by 
the models.

A different future application we plan with CA models extended with our set-up
is the introduction of particles into the models, with the aim to study 
thermal emission, particle acceleration, 
and non-thermal emission. This will allow a much deeper comparison of the
CA models to observations than was possible so far, and this is actually
the most important benefit of the set-up we introduced. 
Such comparisons will allow a new judgment of the adequateness or not
of classical CA models (in their current form) to the problem of solar
flares, beyond the three statistical distributions of the primarily
released energy.
Solar flare CA models which include particle acceleration would represent 
the first global and complete model for solar flares.

\begin{acknowledgements}
We thank K.\ Tsiganis and M.\ Georgoulis for many helpful discussions on 
several issues. We also thank G.\ Einaudi for stimulating discussions on
MHD aspects of flares, and the referee A.L.\ MacKinnon for useful comments.
The work of H.\ Isliker was partly supported by a grant of the Swiss National 
Science Foundation (NF grant nr.\ 8220-046504).
\end{acknowledgements}

\appendix

\section{Temporal evolution of the CA}

\n The temporal evolution of the CA models presented in this article is 
governed by the following rules:

\begin{itemize}
\item[0.] initializing

\item[1.] loading 

\item[2.] scanning: create a list of the unstable sites; if there are none, 
          return to loading (1)

\item[3.] scanning and bursting: redistribute the fields at the unstable sites 
           which are in
           the list created in the scannings 2 or 4

\item[4.] scanning: create a list of the unstable sites. If there are any, go 
          to bursting (3), else return to loading (1)
\end{itemize}

\n The extra scannings 2 and 4 are needed for causality: if a site becomes 
unstable through a burst in the neighbourhood, then it should be redistributed
in the subsequent scan, and not in the same as the primary unstable 
site. The same is true
for the scanning 4, since in the next bursting phase (if any) only those
sites should burst who had become unstable through a burst in their 
neighbourhood during the foregoing time-step.

As a time-step is considered one scanning of the grid, point 3.
The released energy per time-step is the sum of all the energy released
by bursts in this time-step (a burst is considered a single redistribution 
event in 3). We term a flare or avalanche the loop 3,4, from the occurring 
of the first burst in 3 until the activity has died out and one returns
via the scanning 4 to loading (1). The duration of the flare is the number of 
time-steps it lasted, the total flare energy is the sum of all the energies 
released in the duration of the flare, and the peak flux or peak energy is 
the maximum of the energies of all the time-steps of the flare.

\section{Why spline interpolation is particularly adequate: comparison 
to other methods of continuation}

\n We mentioned in Sec.\ 2.2 that other possibilities for continuation of 
the vector-potential besides spline interpolation would be:
a) continuation of $\vec A$ with the help of an equation; b) other kinds of
interpolation, either locally (in a neighbourhood), or globally (through the 
whole grid).
Possibility a) implies that an equation has to be solved in each time-step
(after each loading and after each burst), in the worst case numerically,
with open boundary condition and the $\vec A_{ijk}$ given at the 
grid-sites. This
computational effort might slow down the algorithm of the model 
considerably (and bring it near to the computational effort of MHD
equation integration). Besides that, the problem is what equation to use: to 
make the magnetic field always a potential field (i.e.\ using a 
corresponding equation for the vector-potential $\vec A$) implies that,
from the point of view of MHD, at all times a very 'well-behaved' magnetic 
field resides in the CA, with no tendency towards instabilities, 
which makes it difficult to understand why bursts should occur at all,
since critical quantities such as currents do not become excited.
A better candidate could be expected to be force-freeness, except 
that, possibly, one may be confronted with incompatibility of 
the boundary conditions with the vector-potential values given at the 
grid-sites, i.e.\ existence-problems for solutions eventually arise.

Though definitely possibility a) cannot be ruled out on solid 
grounds, we found it more promising to proceed with possibility b), 
interpolation. 
A guide-line for choosing a particular interpolation method is the reasonable 
demand that the 
interpolation should not introduce wild oscillations in-between grid-sites, 
for we want to assure that the derivatives at the grid sites,
which are very sensitive to such oscillations, are not 'random' values solely
due to the interpolation, but that they reflect more or less directly 
the change of the primary grid-variable from grid-site to grid-site. 
This calls for interpolating functions which are as little curved as possible.

The easiest and fastest way of interpolating would be to perform local 
interpolations around a point and its nearest neighbours 
(e.g.\ using low-order polynomials or trigonometric 
functions of different degrees). This interpolation leads, however, to 
ambiguities for the derivatives:
the derivatives, say at a point $\vec x_{ijk}$, are not the same, if the
used interpolation is centered at $\vec x_{ijk}$, with the ones calculated with 
an interpolation centered at e.g.\ $\vec x_{i+1jk}$. 
In this sense, local interpolation is not self-consistent, 
the derivatives
at a grid-site depend on where the used interpolation is centered.

Finally, we are left with global interpolation through the whole grid. 
Among the candidates are, besides more exotic interpolating functions, 
polynomials of degree equal to the grid size, trigonometric functions
(also in the form of Fourier-transforms), low-order smooth polynomials 
(e.g.\ splines). The first candidate, polynomials of
a high degree $n$ (with $n$ the number of grid points in one direction), we 
reject immediately since it
is notorious for its strong oscillations in-between grid-sites,
mainly towards the edges of the grid. 
We tried the second candidate, trigonometric interpolation, in the form of 
discrete Fourier transform.
Testing this by prescribing analytic functions for $\vec A(\vec x)$ and comparing
the numerical derivatives with the analytic ones,
it turned out that there arise problems with 
representing structures in $\vec A$ as large as the entire grid (the 
wave-number spectrum is too limited), and with structures as short as roughly 
the grid-spacing (different prescribed short structures are taken for the 
same).

Trying cubic spline-interpolation, we found that it does not suffer 
from the problems stated for the other types of interpolation: neither does 
it introduce wild oscillations, unmotivated by the values at the grid-sites,
nor does 
spline interpolation have problems with describing large or small scale 
structures (if a functional form of $\vec A$ is prescribed, then the analytic 
derivatives and the derivatives yielded by the interpolation give very close 
values, in general). 

Moreover, based on results of Sec.\ 3, App.\ C, and Isliker et al.\ (1998), 
there is another
reason why spline-interpolation is particularly adequate to our problem: It 
relates the quantity $d\vec A$ (Eq.\ (2)), which measures 
the stress at a site in the CA model, closely to $\nabla^2 \vec A$, the Laplacian
of $\vec A$ (see App.\ C). The latter is related to the current 
 ($\vec J = -{c\over 4\pi} \nabla^2\vec A + 
 {c\over 4\pi} \nabla(\nabla \vec A)$), 
which, from the point of view of MHD, can be considered as a measure of stress
in the magnetic field configuration. If this relation would not hold, then 
the redistribution rules (Eqs.\ (4) and (5)) of the CA would not be 
interpretable as the 
diffusion process revealed by Isliker et al.\ (1998), and the instability 
criterion (Eq.\ 3) would not be so closely related to the current 
(see Sec.\ 3 and App.\ C).

\subsection{Why in particular differencing is not adequate to calculate 
derivatives in a CA}

\n
We had rejected above (Sec.\ 1, Sec.\ 2.2) the use of difference expressions 
to calculate derivatives, stating that differencing is not in the spirit of CA 
models quite in general, since the nature of CA is truly discrete.
We think it worthwhile to make this argument more concrete and to show
what problems arise if differencing were used:

\n {\sl 1.\ Consistency with the evolution rules:}
Isliker et al.\ (1998) have shown that the classical solar flare CA are not 
just the 
discretized form of a differential equations.
Instead, they describe the time-evolution of a system 
by rules which express the direct transition from a given initial to a 
final state which is the asymptotic solution of a simple 
diffusion equation. 
The time-step corresponds therewith to the 
average time needed
for smallest scale structures (structures as large as a 
neighbourhood) to diffuse, and the grid-size corresponds to the size of these
smallest occurring structures.
Assuming that the CA models were just discretized differential equations would
lead to severe mathematical and physical contradictions and inconsistencies
(continuity for $\Delta h\rightarrow 0$ is violated (with $\Delta h$ the 
grid-size), and negative diffusivities appear). 
Therewith, in order to be consistent with the evolution rules, 
which assume a finite grid-size, one cannot assume for the purpose of 
differentiating this same grid-size to be approximately infinitesimal.

\n{\sl 2.\ Derivatives as difference expressions are not self-consistent:}
There are several equivalent ways to define 
numerical derivatives with the use of difference expressions: there are e.g.\ 
the backward difference 
$\px A_x(\vec x_{ijk})=(A_x(\vec x_{ijk}) - A_x(\vec x_{i-1jk}))/\Delta h$,
and the forward difference 
$\px A_x(\vec x_{ijk})=(A_x(\vec x_{i+1jk}) - A_x(\vec x_{ijk}))/\Delta h$. 
Both should give comparable values in a given 
application, else, in the context of differential equation integration, one 
would have to make the resolution higher. In the case of CA-models, 
we find that the two difference expressions yield values which differ 
substantially from each other:
E.g.\ for an initial loading of the grid with independent random values for 
the $\vec A$-field, the difference between the backward and the forward 
difference 
expression can be as large as the field itself. Such an initial condition 
would of 
course not make sense in the context of partial differential equations, 
in the context of CA, however, it is a reasonable starting configuration, 
and the evolution is unaffected by such an initial loading. Moreover, 
when the CA models we discuss in this article have reached the SOC state,
then the differences between e.g.\ the backward- and 
forward-difference expressions can be as large as 400\%. 
There is no way to reduce this discrepancy, since grid-refinement is 
principally impossible for CA: the evolution is governed by a set of rules, 
and making the grid
spacing smaller by introducing new grid-points in-between the old ones would 
actually just mean to make the grid larger, since the evolution rules remain 
the same, there are no rules for half the grid-spacing.

\section{Relation of $d\vec A$ to $\Delta A$}

The stress measure of LH91, 
$d\vec A_{ijk} = \vec A_{ijk} - {1\over 6} \sum_{n.n.} \vec A_{n.n.}$,
can be related to continuous expressions by representing the values 
of $\vec A_{n.n.}$ as Taylor-series expansions around $\vec x_{ijk}$,
setting the spatial differences to $\Delta h=1$. It turns out that e.g.\ 
\begin{equation}
dA_z = -{1\over 6} \Delta A_z -{1\over 72} (\px^4+\py^4+\pz^4)A_z - ...
\end{equation}
and so on for the other two components. {\it In general,} it is therefore 
{\it not} adequate to consider 
$d\vec A$ to be a good 4th order approximation to $\Delta \vec A$,
since higher order corrections can be large, they depend on the way 
the vector potential is continued in-between grid-sites. If we had, for
instance, chosen global polynomial interpolation instead of 
spline-interpolation, the higher order terms would not be negligible, 
above all towards the edges of the grid, since polynomial interpolation is
known for introducing fluctuations near the edges of the grid. Consequently,
$d\vec A$ would be a bad approximation to $\Delta \vec A$. In order $d\vec A$ to be 
a good approximation to $\Delta \vec A$, interpolation with, for example, 3rd 
order polynomials would be an optimum choice ($d\vec A$ 
would be an exact approximation to $\Delta \vec A$). Thus, 3rd order 
polynomials would be the choice for local interpolation, which, however, is
not applicable, since it introduces discontinuities in $\vec B$ and 
$\vec J$ (see App.\ B). The way out of the dilemma we suggested in this 
article is the use of cubic splines, which provide global interpolation 
with 3rd order polynomials, with $\vec B$ and 
$\vec J$ continuous, and only third order derivatives are discontinuous (this 
is the price of the compromise). For splines then, Eq.\ (C.1) writes as
\begin{eqnarray}
dA_z = -{1\over 6} \Delta A_z - 
    {1\over 36} \Big[  &(&\px^3A_z^+ -\px^3A_z^-) + \cr
                       &(&\py^3A_z^+ -\py^3A_z^-) + \cr
                       &(&\pz^3A_z^+ -\pz^3A_z^-) \Big],
\end{eqnarray}
due to the discontinuities in the 3rd order derivatives (the superscripts
$+$ and $-$ refer to the right and left derivative, respectively). Thus,
in case where the third order right and left derivatives are not too different,
$d\vec A$ is a good approximation to $\Delta \vec A$.

\end{document}